\documentclass{article}

\usepackage{graphicx}
\usepackage{hyperref}
\usepackage{booktabs}
\usepackage{amsmath}

\begin{document}
\title{Geodesic equations and algebro-geometric methods}


\author{Eva Hackmann\footnote{eva.hackmann@zarm.uni-bremen.de}}

\date{ZARM, Universit\"at Bremen, D-28359 Bremen, Germany}

\maketitle

\begin{abstract}
For an investigation of the physical properties of gravitational fields the observation of massive test particles and light is very useful. The characteristic features of a given space-time may be decoded by studying the complete set of all possible geodesic motions. Such a thorough analysis can be accomplished most effectively by using analytical methods to solve the geodesic equation. In this contribution, the use of elliptic functions and their generalizations for solving the geodesic equation in a wide range of well known space-times, which are part of the general Pleba\'nski-Demia\'nski family of solutions, will be presented. In addition, the definition and calculation of observable effects like the perihelion shift will be presented and further applications of the presented methods will be outlined. 
\end{abstract}

\section{Introduction}

The observation of massive particles and light is a very important tool for exploring the features of gravitational fields and also for tests of general relativity. The motion of massive and massless test-particles is described by the geodesic equation, which is a coupled system of ordinary differential equations dependent on the metric of the considered gravitational field. A wide range of exactly known solutions of Einstein's field equations possesses certain symmetries, which allow to decouple the geodesic equations. Here we discuss metrics within the Pleba\'nski-Demia\'nski family of solutions (see \cite{GriffithsPodolsky06}), which is a seven parameter solution with mass, rotation, cosmological constant, electric and magnetic charge, NUT charge, and acceleration and which comprises the Schwarzschild and the Kerr metric as special cases. In this family of solutions the equations of motion considerably simplify due to the separability of the Hamilton-Jacobi equation (for lightlike geodesics and, if the acceleration vanishes, also for timelike geodesics).

Due to this simplification, the analytic solutions for the complete set of geodesics in Schwarzschild space-time was already found in 1931 by Hagihara \cite{Hagihara31} in terms of elliptic functions. With essentially the same methods also the geodesic equations in Kerr-Newman-Taub-NUT space-times and subcases can be handled (and also an additional acceleration for massless particles). For the case of the Kerr metric, this was first done in the equatorial plane (see \cite{Chandrasekhar83} for a review) and later, after the introduction of Mino time \cite{Mino03} which allows to fully decouple the equations of motion, by Fujita and Hikida for bound orbital motion \cite{FujitaHikida09}. For a nonvanishing cosmological constant, the structure of the equations of motions is more complex but can still be solved analytically by using hyperelliptic functions as demonstrated for the Schwarzschild-de Sitter metric \cite{HackmannLaemmerzahl08} and for general axially symmetric space-times \cite{Hackmannetal2009}. Here, we will explain these general methods to analytically solve the geodesic equations in Kerr-Newman-Taub-NUT-de Sitter space-times (and the C-metric for lightlike geodesics). 

For observational purposes also explicit expressions for the deviations of relativistic orbits from the Kepler orbits are of interest. Here we concentrate on the observables for bound orbital motion, namely the periastron shift and the Lense-Thirring effect. However, in a strong gravitational field concepts like the orbital plane and the orbital ellipse are are no longer valid. A fully relativistic treatment of these effects in the Kerr gravitational field was given by Schmidt \cite{Schmidt02} and combined with the Mino time by Drasco and Hughes \cite{DrascoHughes04} as well as Fujita and Hikida \cite{FujitaHikida09}. In this article, we will show how these concepts can be generalized to the above mentioned space-times.

\section{Equations of motion}

The motion of test-particles is described by the geodesic equation
\begin{equation}
\frac{d^2 x^\mu}{ds^2} + \Gamma^{\mu}_{\rho \sigma} \frac{dx^\rho}{ds} \frac{dx^\sigma}{ds} = 0 
\end{equation}
where $\Gamma^{\mu}_{\rho \sigma} = \frac{1}{2} g^{\mu \alpha} (\partial_\rho g_{\sigma\alpha} + \partial_\sigma g_{\rho\alpha} - \partial_\alpha g_{\rho\sigma})$ is the Christoffel symbol and $\mu = 0,1,2,3$. This system of coupled ODE's can be simplified if the underlying space-time has certain symmetries. In particular, for the Pleba\'nski-Demia\'nski solutions (with vanishing acceleration for massive test-particles), there exist four constants of motion which can be used for decoupling: the normalization constant $\epsilon = g_{\mu\nu} \frac{dx^\mu}{ds} \frac{dx^\nu}{ds}$ with $\epsilon=0$ for light and $\epsilon=1$ for massive test-particles, the energy $E$, the angular momentum $L$ in direction of the symmetry axis, and the Carter constant $K$.

For the considered family of solutions in standard Boyer-Lindquist coordinates the radial and latitudinal equations of motion can be reduced to the form \cite{Hackmannetal2009}
\begin{align}\label{generalEOM}
\left( x^i \dfrac{dx}{d\lambda} \right)^2 = P(x;p)
\end{align}
where $x$ is the radius or the (squared) cosine of the latitude, $\lambda$ is an affine parameter, the Mino time \cite{Mino03}, $P$ is a polynomial in $x$ of degree $2g+1$ or $2g+2$, $0\leq i<g$ is an integer, and $p=\{p_1,\ldots,p_n\}$ is a set of parameters of the space-time and the test-particle. For example, in Kerr space-time we get in geometrized units ($G=1$, $c=1$) \cite{Chandrasekhar83, Mino03}
\begin{align}
\left(\frac{dr}{d\lambda}\right)^2 & = ((r^2+a^2)E-aL)^2-(r^2+a^2-2r)(\epsilon r^2+K)=:R(r)\,,\\
\left(\frac{d\xi}{d\lambda}\right)^2 & = 4\xi [(1-\xi) (K-\epsilon a^2\xi) + (aE(1-\xi)-L)^2]\,,
\end{align}
where $\xi=\cos^2\theta$ and all quantities are normalized such that they are dimensionless.
%

\section{Algebro-geometric methods}

\begin{figure}
\centering
\includegraphics[width=0.27\textwidth]{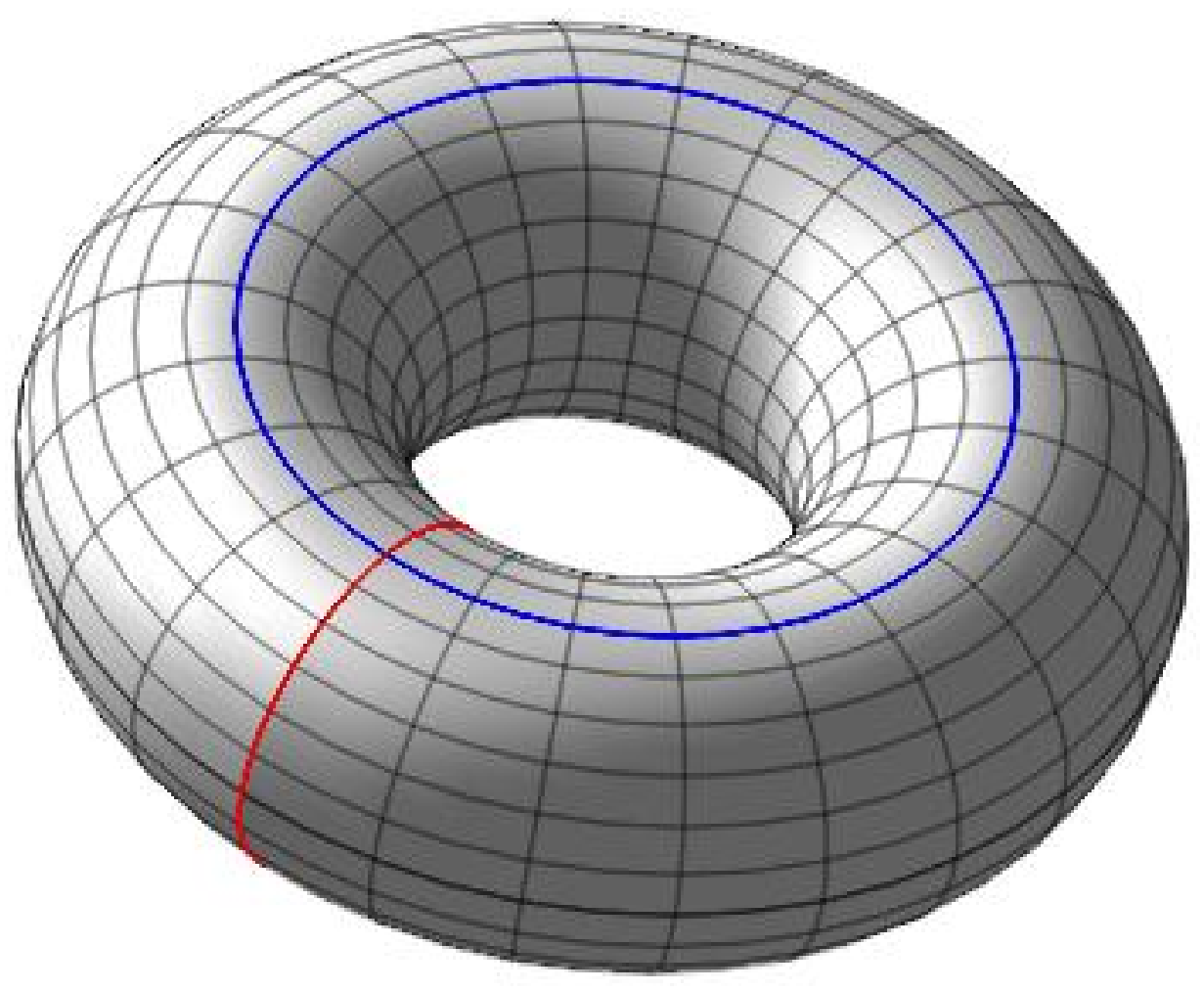}\qquad
\includegraphics[width=0.32\textwidth]{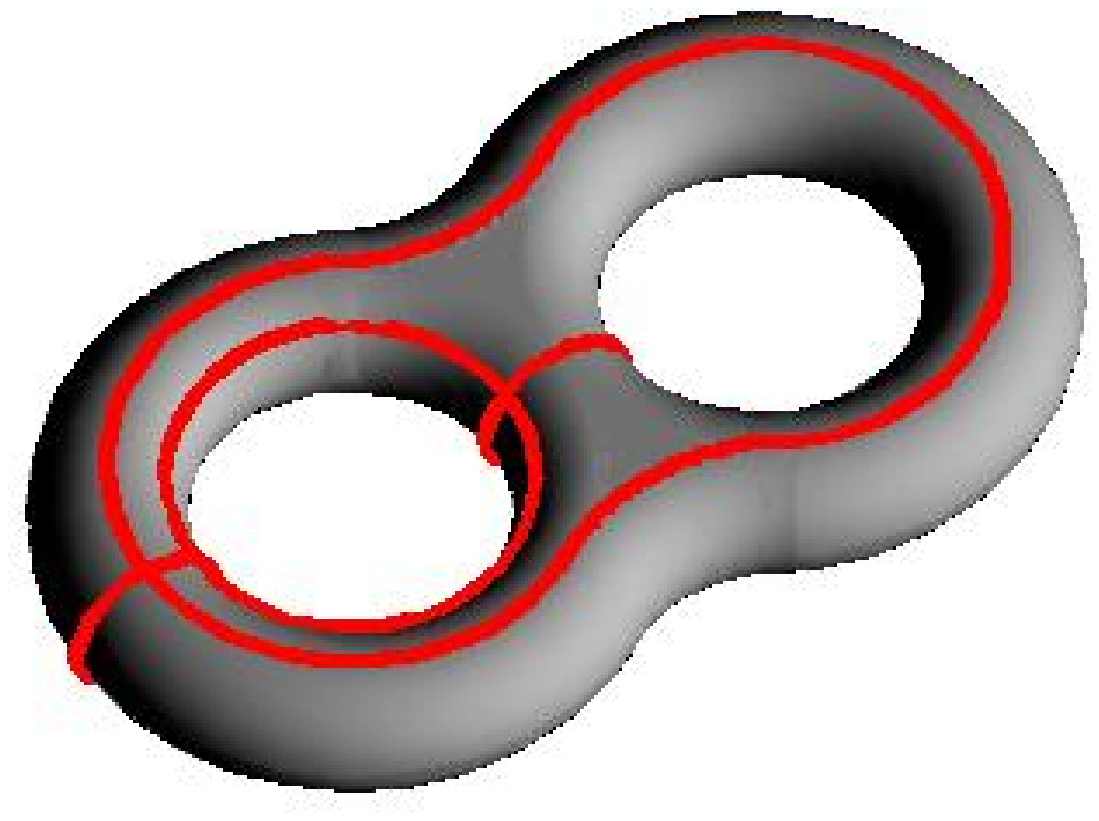}
\caption{Riemann surface of a genus one curve (left) and a genus two curve (right) with $2$ or $4$ independent paths.}
\label{Fig:Riemannsurface}
\end{figure}

The equation of motion \eqref{generalEOM} should be solved for $x(\lambda)$, i.e.~we consider the inversion problem
\begin{align}
\int_{x_0}^x \frac{x^i dx}{\sqrt{P(x;p)}} = \lambda-\lambda_0\,,
\end{align}
where $x(\lambda_0)=x_0$ are initial values. The solution of this problem should be independent from the chosen integration path. This implies that for a closed path $\gamma$ with $\omega := \oint_\gamma \frac{x^i dx}{\sqrt{P(x;p)}} \neq 0$
the solution $x(\lambda)$ has to have the period $\omega$, $\int_{x_0}^x \frac{x^i dx}{\sqrt{P(x;p)}} = \lambda-\lambda_0-\omega$.
This can be taken into account automatically if the equation \eqref{generalEOM} is considered as an algebraic curve
\begin{align}\label{generalcurve}
w^2 = P(x;p)\,,
\end{align}
where $w=x^i\frac{dx}{d\lambda}$. For the considered space-times one of the following two situations occur
\begin{itemize}
\item $P$ is of order $3$ or $4$: then \eqref{generalcurve} is an elliptic curve of genus $1$,
\item $P$ is of order $5$ or $6$: then \eqref{generalcurve} is an hyperelliptic curve of genus $2$.
\end{itemize}
Topologically, (hyper-)elliptic curves can be considered as Riemann surfaces. The genus $g$ corresponds to the number of 'holes' in the Riemann surface, see Figure \ref{Fig:Riemannsurface}. If there are $g$ holes this implies that there are $2g$ independent closed integration paths whose integrals do not vanish and, therefore, the solution $x(\lambda)$ needs to have $2g$ independent periods \cite{FarkasKra1992}.

In the case of an elliptic curve \eqref{generalcurve} can be reduced to the Weierstrass form by a rational substitution,
\begin{align}\label{Weierstrassform}
\tilde{w}^2 = 4\tilde{x}^3-g_2\tilde{x}-g_3\,.
\end{align}
In this standard form, $\tilde{w}$ and $\tilde{x}$ are parametrized by the Weierstrass elliptic function and its derivative, $\tilde{w}=\wp'(z)$ and $\tilde{x}=\wp(z)$. In the above example of Kerr space-time, the substitution for the radial equation of motion for a timelike geodesic is $r=\frac{a_3}{4x-\frac{a_2}{3}}+r_K$, where $R(r_K)=0$ and $a_j=\frac{1}{(4-j)!} \frac{d^{(4-j)}R}{dr^{(4-j)}}(r_K)$. The resulting equation is given by \eqref{Weierstrassform} with
\begin{align}
g_2 & = \frac{1}{4} \left( \frac{1}{3} a_2^2 -a_1a_2 \right), \quad g_3 = \frac{1}{16} \left( \frac{1}{3} a_1a_2a_3 - \frac{2}{27}a_2^3 -a_0a_3^2 \right) \,.
\end{align}
The analytical solution for the timelike radial equation in Kerr space-time is then given by
\begin{align}
r(\lambda) & = \frac{a_3}{4\wp(\lambda-c)-\frac{a_2}{3}} + r_K\,,
\end{align}
where $c=c(r_0,\lambda_0)$ is a constant which depends only on the initial conditions.

With a complete analogous procedure all equations of motions which reduce to elliptic curves may be solved. This includes geodesics in Schwarzschild, Reissner-Nordstr\"om, Kerr-Newman, and Taub-NUT space-times as well as the C-metric (see e.g.~\cite{Bicaketal1989}) for massless particles.

The situation gets more complicated if the equation of motion is described by a hyperelliptic curve of genus two. This is due to the fact that the solution has to have four independent periods, which is impossible for functions in a single complex variable. Therefore, is is necessary to consider a function in $g=2$ complex variables. However, as we have only one degree of freedom, we need to restrict the space of variables again to a one-dimensional submanifold. This is done by considering the equation of motion as part of the Jacobi inversion problem for $g=2$, which is to solve the system of equations
\begin{equation}
\begin{aligned}
w_1 & = \int_{\infty}^{x_1} \frac{dx}{\sqrt{P(x)}} + \int_{\infty}^{x_2} \frac{dx}{\sqrt{P(x)}}\,,\\
w_2 & = \int_{\infty}^{x_1} \frac{xdx}{\sqrt{P(x)}} + \int_{\infty}^{x_2} \frac{xdx}{\sqrt{P(x)}}\,,
\end{aligned}
\end{equation}
for $x_1$, $x_2$ as functions of $w_1$, $w_2$. If $P$ is transformed to a standard form $P(x) = 4x^5 + \sum_{i=0}^4 a_ix^i$ the solution to this problem is known in terms of generalized $\wp$-functions, $x_1x_2 = \wp_{12}(w_1,w_2)$, $x_1+x_2=\wp_{22}(w_1,w_2)$ \cite{FarkasKra1992}. Here $\wp_{ij}(w_1,w_2) = - \frac{\partial}{\partial w_i} \frac{\partial}{\partial w_j} \log \sigma(w_1,w_2)$ with the generalized $\sigma$ function. By letting $x_2$ go to infinity, we can get rid of the second integral on the right hand side and simultaneously restrict $(w_1,w_2)$ to the one-dimensional sigma divisor, i.e.~the set of zeros of the two-dimensional $\sigma$-function. The solution for $x=x_1$ is then given by
\begin{align}
x = - \frac{\sigma_1}{\sigma_2}(w_1,w_2)\,, \qquad \text{with} \quad \sigma(w_1,w_2)=0\,,
\end{align}
where $\sigma_i$ denotes derivative of $\sigma$ w.r.t.~the i-th variable. For example, in the case of Schwarzschild-de Sitter space-time, the solution for the radius $r$ in terms of the angle $\varphi$ is given by \cite{HackmannLaemmerzahl08}
\begin{align}
r(\varphi) = - M \frac{\sigma_2}{\sigma_1}(f(\varphi),\varphi)\,, \qquad \text{with} \quad \sigma(f(\varphi),\varphi)=0\,.
\end{align}
This solution method can be applied to all geodesic equations which reduce to hyperelliptic curves. This comprises the radial and latitudinal equations of motion in Kerr-de Sitter space-time as well as all the de Sitter-versions of Kerr-Newman-Taub-NUT space-times and subcases \cite{Hackmannetal2009}. For lightlike geodesics also the C-metric with an additional cosmological constant can be treated this way.

\section{Observables}
For the bound orbital motion of massive test-particles there are two main gravitational effects
: the periastron shift $\Delta_{\rm P}$ and the Lense-Thirring effect $\Delta_{\rm LT}$. The first is a precession of the orbital ellipse within its orbital plane and the latter a precession of the orbital plane itself.\footnote{In the original paper \cite{LenseThirring18} a combined effect was computed: the timely variation of the ascending node and the argument of periapsis. With 'Lense-Thirring effect' we refer only to the first correction.} In this picture, the periastron shift is defined as the angle between two consecutive periastrons and the Lense-Thirring effect as the angle between two consecutive minimal latitudes. Mathematically this means that they are given by the difference between the period of $r(\varphi)$ or $\theta(\varphi)$, respectively, and $2\pi$. If $2\omega_r$ denotes the period of $r(\varphi)$ and $2\omega_\theta$ the period of $\theta(\varphi)$ then
\begin{align}
\Delta_{\rm P} & = 2\omega_r - 2 \,{ \rm sign }(L)\, \pi\,, \qquad \Delta_{\rm LT} = 2\omega_\theta - 2 \,{ \rm sign }(L)\, \pi\,,
\end{align}
where the sign of $L$ is included to distinguish between prograde and retrograde motion.


In Schwarzschild space-time, where $r(\varphi)$ is directly known, the exact analytical expression for the periastron shift is given by
\begin{align}
\Delta_{\rm P} = \frac{4LK(k)}{\sqrt{(E^2-1)r_{\rm p}(r_{\rm a}-r_1)}} -2\pi\,,
\end{align}
where $K(k)$ is the complete elliptic integral of the first kind,
\begin{align}\label{ellipticfirstkind}
K(k)=\int_0^1 \frac{dx}{\sqrt{(1-x^2)(1-k^2x^2)}}\,,
\end{align}
which is implemented in standard mathematical software. Here $k^2 = \frac{r_1(r_{\rm a}-r_{\rm p})}{r_{\rm p}(r_{\rm a}-r_1)}$ with the zeros $0<r_1<r_{\rm p}<r_{\rm a}$ of $R(r)=\left(\frac{dr}{d\varphi}\right)^2$.

The expression for the perihelion shift can be generalized to axially symmetric space-times, where only $r(\lambda)$ and $\varphi(\lambda)$ is known. Writing $\varphi(\lambda)$ as a part linear in $\lambda$, given in form of an infinite Mino time average $\Upsilon_\varphi$, plus oscillatory deviations \cite{Schmidt02,FujitaHikida09}, we may use $\lambda(\varphi)=\Upsilon_\varphi^{-1} \varphi$. Then the period of $r(\varphi)$ is given by
\begin{align}
r(\lambda(\varphi+2\varLambda_r\Upsilon_\varphi)) = r(\Upsilon_\varphi^{-1} \varphi + 2\varLambda_r) = r(\Upsilon_\varphi^{-1} \varphi) = r(\lambda(\varphi))\,,
\end{align}
where $2\varLambda_r$ is the period of $r(\lambda)$. Accordingly, $2\omega_r=2\varLambda_r\Upsilon_\varphi$ and the perihelion shift can be written as
\begin{align}
\Delta_{\rm P} = 2\varLambda_r\Upsilon_\varphi - 2\, {\rm sign }(L) \, \pi\,.
\end{align}

In a totally analogous way the Lense-Thirring effect can be found. If $2\varLambda_\theta$ denotes the period of $\theta(\lambda)$ we may write again $\theta(\lambda(\varphi+2\varLambda_\theta\Upsilon_\varphi))=\theta(\lambda(\varphi))$ and, therefore,
\begin{align}
\Delta_{\rm LT} = 2\varLambda_\theta \Upsilon_\varphi -  2 \,{\rm sign }(L)\, \pi\,.
\end{align}

In the case of Kerr-Newman-Taub-NUT-de Sitter space-times, $\varLambda_{r,\theta}$ and $\Upsilon_\varphi$ are given in terms of (hyper-)elliptic integrals. For the elliptic case, they can be rewritten in terms of the three standard Jacobian elliptic integrals $K(k)$, $E(k)$, and $\Pi(n,k)$. For the case of genus two or higher, to our knowledge such a standard form does not exist. A possible generalization of the first Jacobian elliptic integral would be
\begin{align}
K_{\vec{A}}(\vec{k}) & = \int_0^1 \frac{\sum_{i=1}^{g} A_i t^{i-1} dt}{\sqrt{t(1-t)\prod_{i=1}^{2g-1}(1-k_i^2t)}}\,, \label{Completehyperelliptic}
\end{align}
where $\vec{A}$ is a vector of length $g$, which reflects the fact that there are $g$ independent differentials of the first kind, and $\vec{k}$ is of length $2g-1$. 

For example, in terms of these integrals the perihelion shift for Schwarzschild-de Sitter space-time is given by
\begin{align}
\Delta_{\rm P} = \frac{r_4 K_{\left(\frac{1}{r_4},\frac{(r_4-r_3)}{r_3r_4}\right)}(k_1,k_2,k_3)}{\sqrt{L^2\Lambda r_3(r_4-r_0)(r_4-r_2)(r_5-r_4)}} -2\pi\,,
\end{align}
where $r_0<0<r_1<r_2<r_3<r_4<r_5$ are the zeros of the defining polynomial with $r_3=r_{\rm min}$ and $r_4=r_{\rm max}$ for small positive $\Lambda$ and
\begin{align}
k_1^2 & = \frac{r_0(r_4-r_3)}{r_3(r_4-r_0)}\,, \quad k_2^2 = \frac{r_2(r_4-r_3)}{r_3(r_4-r_2)}\,, \quad k_3^2 = -\frac{r_5(r_4-r_3)}{r_3(r_5-r_4)}\,.\end{align}

\section{Outlook}

The methods presented here are powerful tools for the analytical integration of the geodesic equation in a wide range of space-times. Beside the space-times we focused on here, geodesics in higher-dimensional spherically symmetric space-times \cite{Hackmannetal2008, Enolskiietal2011} and the Meyers-Perry space-time \cite{KagramanovaReimers2012} can be treated. It may also be possible to extend these methods to the equations of motion in space-times with given multipole moments like the Erez-Rosen space-time.

The hyperelliptic curves which we used here to represent the equations of motions are a special case of abelian curves, which allow for higher orders of $w$ in \eqref{generalcurve} and mixed terms. These more general curves appear for example in Ho\v{r}ava-Lifshitz and Gauss-Bonnet gravity, which may be represented by quartic curves of the form $(w-P(x))^2=Q(x)$. A generalization of the presented methods to these cases is in preparation.

Analogously to the analytic expressions presented here for observables of bound orbital motion also the bending of light and the gravitational time delay may be considered. Linked to that, we plan the development of an analytical timing formula for pulsars orbiting a black hole.

Concerning the numerical calculation of the analytical expressions, the complete elliptic integrals can very efficiently be computed by using the arithmetic geometric mean. This can be generalized to genus two hyperelliptic integrals, see \cite{BostMestre88,Bradenetal2011}. 

\bibliographystyle{unsrt}
\bibliography{ae100prg_Hackmann}

\begin{thebibliography}{10}

\bibitem{GriffithsPodolsky06}
J.B. Griffiths and J.~Podolsky.
\newblock {A new look at the Plebanski-Demianski family of solutions}.
\newblock {\em Int. J. Mod. Phys.}, 15:335, 2006.

\bibitem{Hagihara31}
Y.~Hagihara.
\newblock Theory of relativistic trajectories in a gravitational field of
  {S}chwarzschild.
\newblock {\em Japan. J. Astron. Geophys.}, 8:67, 1931.

\bibitem{Chandrasekhar83}
S.~Chandrasekhar.
\newblock {\em The Mathematical Theory of Black Holes}.
\newblock Oxford University Press, Oxford, 1983.

\bibitem{Mino03}
Y.~Mino.
\newblock Perturbative approach to an orbital evolution around a supermassive
  black hole.
\newblock {\em Phys. Rev. D}, 67:084027, 2003.

\bibitem{FujitaHikida09}
R.~Fujita and W.~Hikida.
\newblock Analytical solutions of bound timelike geodesic orbits in {K}err
  spacetime.
\newblock {\em Class. Quantum Grav.}, 26:135002, 2009.

\bibitem{HackmannLaemmerzahl08}
E.~Hackmann and C.~L{\"a}mmerzahl.
\newblock {Complete analytic solution of the geodesic equation in
  Schwarzschild--(anti) de Sitter space--times}.
\newblock {\em Phys. Rev. Lett.}, 100:171101, 2008.

\bibitem{Hackmannetal2009}
E.~et~al Hackmann.
\newblock Analytic solutions of the geodesic equation in axially symmetric
  space-times.
\newblock {\em Europhys. Lett.}, 88:30008, 2009.

\bibitem{Schmidt02}
W.~Schmidt.
\newblock Celestial mechanics in {K}err spacetime.
\newblock {\em Class. Quantum Grav.}, 19:2743, 2002.

\bibitem{DrascoHughes04}
S.~Drasco and S.A. Hughes.
\newblock Rotating black hole orbit functionals in the frequency domain.
\newblock {\em Phys. Rev. D}, 69:044015, 2004.

\bibitem{FarkasKra1992}
H.M. Farkas and I.~Kra.
\newblock {\em Riemann Surfaces}.
\newblock Graduate Texts in Mathematics. Springer-Verlag, 1992.

\bibitem{Bicaketal1989}
J.~Bi\v{c}\'{a}k and B.~Schmidt.
\newblock Asymptotically flat radiative space-times with boost-rotation
  symmetry: The general structure.
\newblock {\em Phys. Rev. D}, 40:1827, 1989.

\bibitem{LenseThirring18}
J.~Lense and H.~Thirring.
\newblock {\"Uber den Einflu{\ss} der Eigenrotation der Zentralk\"orper auf die
  Bewegung der Planeten und Monde nach der Einsteinschen Gravitations-theorie}.
\newblock {\em Phys. Zeitschrift}, 19:156, 1918.

\bibitem{Hackmannetal2008}
E.~et~al Hackmann.
\newblock Analytic solutions of the geodesic equation in higher dimensional
  static spherically symmetric spacetimes.
\newblock {\em Phys. Rev. D}, 78:124018, 2008.

\bibitem{Enolskiietal2011}
V.Z. et~al Enolskii.
\newblock Inversion of hyperelliptic integrals of arbitrary genus with
  application to particle motion in general relativity.
\newblock {\em J. Geom. Phys.}, 61:899, 2011.

\bibitem{KagramanovaReimers2012}
V.~Kagramanova and S.~Reimers.
\newblock Analytic treatment of geodesics in five-dimensional myers-perry
  space-times.

\bibitem{BostMestre88}
J.-B. Bost and J.F. Mestre.
\newblock Moyenne arithm\'{e}tico-g\'{e}om\'{e}trique et p\'{e}riodes des
  courbes de genre 1 et 2.
\newblock {\em Gaz.Math.Soc.France}, 38:36, 1988.

\bibitem{Bradenetal2011}
H.W. Braden, A.~D'Avanzo, and V.Z. Enolski.
\newblock On charge-3 cyclic monopoles.
\newblock {\em Nonlinearity}, 24:643, 2011.

\end{thebibliography}

\end{document}